%% file: dihiggs_draft_jhep.tex
\documentclass[a4paper,11pt]{article}
\pdfoutput=1 % if your are submitting a pdflatex (i.e. if you have
\usepackage{jheppub} % for details on the use of the package, please
\usepackage[T1]{fontenc} % if needed
\usepackage{multirow}
\usepackage{graphicx}
\usepackage{xspace} %load xspace
\usepackage{lineno}
\usepackage{float}
\usepackage[caption = false]{subfig}

\usepackage{bookmark} % helps booksmarks look better in PDF
\hypersetup{colorlinks=true,linkcolor=blue, breaklinks} %internal links in blue, citations in green
\usepackage[all]{hypcap}

\include{latex/shortcuts} %load list of shortcuts shortcuts.tex

\title{\boldmath Benchmarking Machine Learning Techniques with Di-Higgs Production at the LHC}

\author[a]{B. Tannenwald}
\author[a]{C. Neu,}
\author[a]{A. Li,}
\author[a]{G. Buehlmann,}
\author[a]{A. Cuddeback,}
\author[a]{L. Hatfield,}
\author[a]{R. Parvatam,}
\author[a]{C. Thompson}

\affiliation[a]{University of Virginia, 248 McCormick Road, Charlottesville, VA, USA}

\emailAdd{benjamin.tannenwald@cern.ch}
\input{abstract.tex}

\begin{document} 
%\linenumbers # option for including line numbers in draft-mode
\maketitle
\flushbottom

%%% Main body of paper
\input{intro.tex}

%\input{defaultContent_1.tex}
%\input{defaultContent_2.tex}

\input{dihiggsPhys/dihiggs_physics.tex}

\input{supervised_ml.tex}

\input{semi-supervised_ml.tex}

\input{results.tex}

\input{conclusions.tex}

% Extra stuff
%\input{acknowledgements.tex}

%\backmatter
% We use BIBTeX for the bibliography---you don't have to
%\input{bibliography.tex}

\nocite{*} % To display all refs, even uncited refs (useful when editting)
%\bibliography

\cleardoublepage
\phantomsection

%\addcontentsline{toc}{chapter}{Bibliography}
%\bibliographystyle{ieeetr}
\bibliographystyle{JHEP}
\bibliography{bibliography}

%\bibliographystyle{unsrt} % use your favorite BIBTeX style

% APPENDICES IF NEEDED
%\input{appendices/appendixA.tex}

\end{document}

%% file: latex/shortcuts.tex
\newcommand*{\pt}{\ensuremath{p_{\text{T}}}\xspace}

\newcommand*{\TeV}{\ifmmode {\mathrm{\ Te\kern -0.1em V}}\else
                   \textrm{Te\kern -0.1em V}\fi}%
\newcommand*{\GeV}{\ifmmode {\mathrm{\ Ge\kern -0.1em V}}\else
                   \textrm{Ge\kern -0.1em V}\fi}%
\newcommand*{\MeV}{\ifmmode {\mathrm{\ Me\kern -0.1em V}}\else
                   \textrm{Me\kern -0.1em V}\fi}%
\newcommand*{\keV}{\ifmmode {\mathrm{\ ke\kern -0.1em V}}\else
                   \textrm{ke\kern -0.1em V}\fi}%
\newcommand*{\eV}{\ifmmode  {\mathrm{\ e\kern -0.1em V}}\else
                   \textrm{e\kern -0.1em V}\fi}%

%% file: abstract.tex
\abstract{Many domains of high energy physics analysis are starting to explore machine learning techniques. Powerful methods can be used to identify and measure rare processes from previously insurmountable backgrounds. One of the most profound Standard Model signatures still to be discovered at the LHC is the pair production of Higgs bosons through the Higgs self-coupling. The small cross section of this process makes detection very difficult even for the decay channel with the largest branching fraction ($hh\rightarrow b\bar{b}b\bar{b}$). This paper benchmarks a variety of approaches (boosted decision trees, various neural network architectures, semi-supervised algorithms) against one another to catalog a few of the various techniques available to high energy physicists as the era of the HL-LHC approaches.}

%% file: intro.tex
\section{Introduction}
\label{sec:intro}

The use of machine learning (ML) techniques in high energy particle physics has rapidly expanded in the last few decades. The proliferation of methods and applications has touched nearly every segment of analysis and reconstruction~\cite{albertsson2018machine} and will be vital in understanding the full dataset of the Large Hadron Collider (LHC) and data from future colliders.

Common approaches involve using linear techniques like decision trees and non-linear approaches like neural networks. These techniques are then used to reconstruct objects like leptons and jets, to tag b-quarks and boosted decays, and to classify different processes. Many models depend on kinematic input features physicists have traditionally used; other architectures rely on emergent features produced in a more abstract phase-space. With so many usable machine-learning options available, interesting questions arise about what types of information are best to feed to our networks. The information that is best for physicists to learn from might not be optimal for sophisticated computing algorithms.
%You can even think of fun semi-supervised approaches like clustering and maybe some stuff with autoencoders. To be totally honest, I'm not really sure if the autoencoder stuff is relevant for this paper, but it's really interesting and is maybe cool for unexpected BSM signatures? Could be a cool side-study.

The goal of this paper is to explore a wide range of current ML techniques that can be used to identify di-Higgs production at the High Luminosity LHC (HL-LHC). Observing di-Higgs production is necessary to measure the self-coupling of the Higgs boson and fully understand the nature of electroweak symmetry breaking. The difficulty in measuring Higgs pair production lies in the tiny cross-section of even the largest branching fraction ($hh\rightarrow b\bar{b}b\bar{b}$) and the relative abundance of similarly reconstructed QCD events.

This paper is organized as follows: Section 2 deals with the physics relevant for di-Higgs production and the QCD background. Sections 3 and 4 summarize the various ML methods tested. The maximum significance
\begin{equation}
  \sigma = \frac{N_{\textrm{signal}}}{\sqrt{N_{\textrm{background}}}}
\end{equation}
is provided for each method along with event yields normalized to the HL-LHC design integrated luminosity of 3000 fb$^{-1}$. Section 5 compares the results from the various approaches.

%% file: dihiggsPhys/dihiggs_physics.tex
%Dihiggs physics is cool. Also complicated but not that complicated. Mostly just rare. Show some diagrams. Talk about rates.
\section{Di-Higgs Physics}
\subsection{Higgs Pair Production}
\label{sec:physics}

\begin{figure}[!h] 
\begin{center}
\includegraphics*[width=0.75\textwidth] {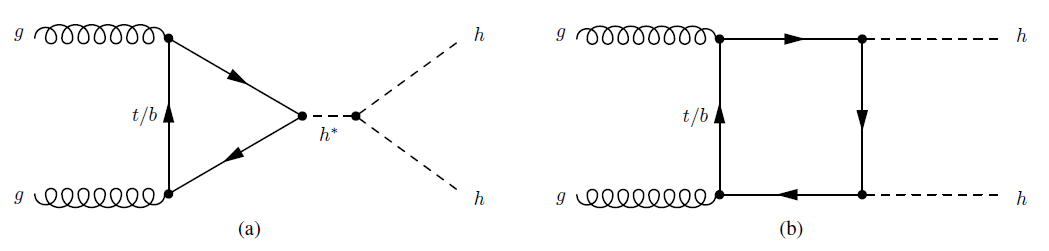}
\caption{Leading order Feynman diagrams for non-resonant production of Higgs
  boson pairs in the Standard Model through (a) the Higgs boson self-coupling
  and (b) the Higgs-fermion Yukawa interaction.} 
  \label{fig:nr_hh_production}
\end{center}
\end{figure}

The Higgs boson is an essential part of the Standard Model (SM) of particle physics and is a product of the mechanism responsible for electroweak symmetry breaking. Along with the interactions of the Higgs boson to the other SM particles, an additional Higgs self-coupling interaction is predicted at tree-level by the SM. This mechanism contributes to non-resonant Higgs boson pair production together with quark-loop contributions via Yukawa-type interactions. Figure~\ref{fig:nr_hh_production} shows the two leading order diagrams of non-resonant Higgs boson pair production. Since the production cross section for Higgs boson pair production is extremely small within the SM~\cite{deFlorian:2016spz}, 

\begin{equation*}
\sigma_{hh}\text{ (14 TeV)} = 39.6 \pm 2.7 \text{ fb},
\end{equation*}
any significant deviation would indicate the presence of new physics.

Many extensions of the SM predict the existence of additional scalar bosons which may have mass larger than twice the Higgs mass and can decay into a Higgs boson pair. Searching for resonances in the $hh$ mass spectrum can help us discover or limit exotic models which predict the presence of such particles. More importantly, measuring the SM di-Higgs cross-section (or placing limits on its magnitude) allows us to probe the self-coupling of the Higgs field and better understand the mechanism behind electroweak symmetry breaking.

The following work is focused on techniques for distinguishing non-resonant (SM-like) Higgs boson pair production where both Higgs bosons decay via $h \to b \bar{b}$. Using the $4b$ decay mode provides the largest possible amount of signal events but requires powerful background reduction techniques due to the large production cross-section of fully hadronic QCD processes. All results are quoted for simulated events produced by $pp$ collisions with a center-of-mass energy of 14 TeV and scaled to the HL-LHC design integrated luminosity of 3000 fb$^{-1}$. Simulated samples were produced using Madgraph v2.7.0~\cite{Alwall:2014hca} in a ROOT v6.12/04~\cite{Brun:1997pa} environment. Events were then showered using Pythia v8.2.44~\cite{Sj_strand_2015} and reconstructed with Delphes v3.0~\cite{de_Favereau_2014} using the v2 approximation of the upgraded Phase-II CMS detector.

Both the signal and background samples were generated with minimal pileup addition sampled from a Poisson distribution with an expectation value of zero additional vertices. An additional generator-level cut requiring total hadronic energy greater than 300 GeV was applied when generating background QCD events. All code used to set up the generation environment and produce events is publicly available~\cite{github}. A summary of the sample generation details is shown in Table~\ref{tab:samples}. The effective cross-sections in Table~\ref{tab:samples} are used for all event yield calculations in subsequent sections.

\begin{table}[ht!]
 \label{tab:samples}
\centering
  %\begin{center}
    \begin{tabular}{|l|c|c|c|} % <-- Alignments: 1st column left, 2nd middle and 3rd right, with vertical lines in between
      \hline\hline
      Name & Process & $\sigma_{\textrm{eff}}$ [fb] & $N_{\textrm{events}}$ \\
      \hline
      Di-Higgs & $p p \rightarrow h h$, $h \rightarrow b \bar{b}$ & $12.4 \pm 0.1$ & 1$\cdot 10^6$ \\
      QCD     & $p p \rightarrow b \bar{b} b \bar{b}$ & $441900 \pm 2300$ & 4$\cdot 10^6$ \\
      \hline\hline
    \end{tabular}
\caption{Madgraph processes, effective cross-sections, and number of generated events for the signal and background samples used in this paper. The effective cross-sections differ from the total theoretical cross-sections due to branching fractions and generation-level cuts on hadronic activity.}
\end{table}

The assumption of zero additional background interactions per event is an unrealistic approximation of the HL-LHC environment where an average of two hundred additional pileup interactions is expected for each collision. This paper is an initial study aimed at gauging the power of various techniques in a simplified environment. Future studies will extend the approaches explored here to a realistic HL-LHC regime.

\subsection{Event Reconstruction}
\label{sec:eventReco}
\begin{figure}[ht!]
  \begin{center}
  \subfloat[]{\includegraphics[width = 2.5in]{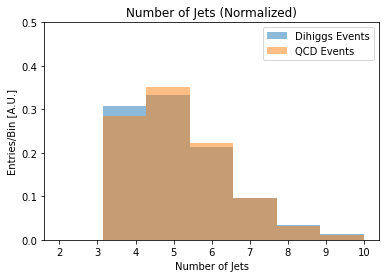}}
  \subfloat[]{\includegraphics[width = 2.5in]{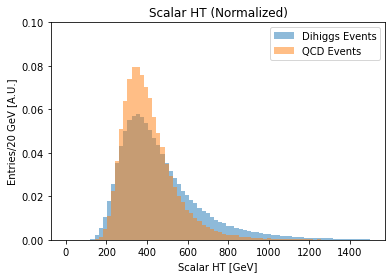}}\\
  \subfloat[]{\includegraphics[width = 2.5in]{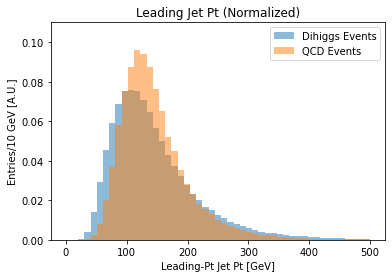}}
  \subfloat[]{\includegraphics[width = 2.5in]{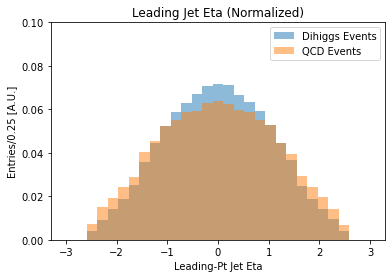}}
  \caption{Sample of event information and leading jet kinematics from QCD and di-Higgs simulation. Distributions are normalized to the same area to compare shapes.}
  \label{fig:jetInfo}
  \end{center}
\end{figure}

The first step in reconstructing the 4$b$ system is to reconstruct and select b-jet candidates. Jets are clustered using an anti-$k_T$ algorithm with a radius of R=0.4. To be selected for use in event reconstruction, a jet must have $\pt > 20$ GeV and $\mid\eta\mid < 2.5$. Delphes uses an internal $b$-tagging efficiency parameterization to predict whether jets are tagged. An event is only fully reconstructed if at least 4 jets are $b$-tagged (unless otherwise specified). The properties and kinematics of selected jets are shown in Figure~\ref{fig:jetInfo}. This strict requirement of having at least 4 $b$-tags in an event helps to reduce contributions from QCD and resolve some of the combinatoric ambiguity in event reconstruction.

\begin{figure}[ht!]
  \begin{center}
  \subfloat[]{\includegraphics[width = 2in]{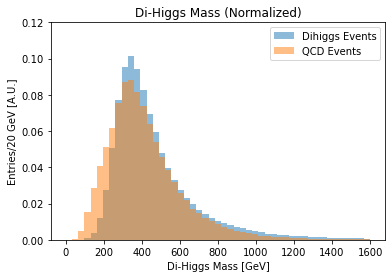}} 
  \subfloat[]{\includegraphics[width = 2in]{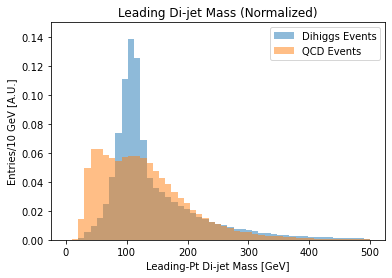}}
  \subfloat[]{\includegraphics[width = 2in]{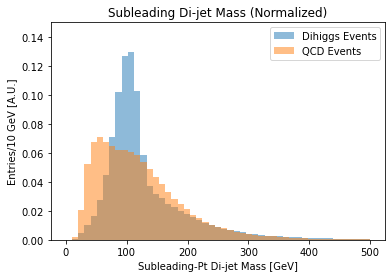}}\\
  \subfloat[]{\includegraphics[width = 2in]{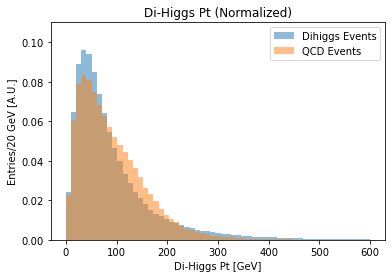}}
  \subfloat[]{\includegraphics[width = 2in]{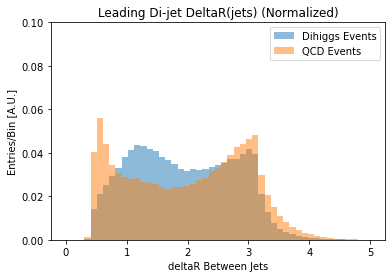}}
  \subfloat[]{\includegraphics[width = 2in]{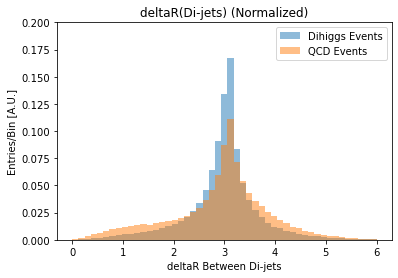}} 
  \caption{Sample of reconstructed kinematics of the di-Higgs system in QCD and di-Higgs simulation. Distributions are normalized to the same area to compare shapes.}
  \label{fig:recoInfo}
  \end{center}
\end{figure}

Once events with at least 4 $b$-tags are selected, there is a choice about how to reconstruct the di-Higgs system. Several reconstruction methods were tested for pairing b-jets to find an optimal algorithm for correctly pairing Higgs boson constituents. Two algorithms were selected for use in the following sections: the first iterates through all selected jets in an event and returns the two pairs with closest di-jet masses to one another. The second returns the two jet pairs that yield two Higgs candidates with masses closest to the known Higgs boson mass by minimizing the quantity 

\begin{equation*}
\chi^2 = \left ( \frac{m_1 - m_h}{\sigma_h} \right ) ^2 + \left ( \frac{m_2 - m_h}{\sigma_h} \right ) ^2
\end{equation*}
where $m_{1,2}$ are the candidate di-jet masses, $m_h=125$ GeV is Higgs boson mass, and $\sigma_h =10$ GeV is the expected experimental resolution of the Higgs boson in the $b\bar{b}$ decay channel. Unless otherwise specified, the method that selects di-jets with masses closest to each other is used when training algorithms that require reconstructed events. Fig~\ref{fig:recoInfo} shows a selection of distributions describing the di-Higgs system using this reconstruction algorithm.

%\begin{figure}[ht!]
%  \begin{center}
%  \subfloat[]{\includegraphics[width = 2.5in]{dihiggsPhys/figures/reconstructedQuantities/mh_pairJetsUsingSmallestDeltaR}} 
%  \subfloat[]{\includegraphics[width = 2.5in]{dihiggsPhys/figures/reconstructedQuantities/mh_pairJetsClosestToHiggs}}\\
%  \caption{Leading dijet mass using different reconstruction algorithms. Left plot shows mass when minimizing dijet angular separation, right plot shows mass when minimizing different between dijet mass and a Higgs mass of 125 GeV ('closestDijetMassToHiggs'). [PLACEHOLDER PLOTS]}
%  \label{fig:compareReco}
%  \end{center}
%\end{figure}

Reconstructed variables include the masses and momentum of the two- and four-body Higgs candidates as well as the angular separations between the two Higgs candidates and their constituent jets. Additional event-level variables like the number of selected jets, the number of b-tagged jets, and the missing transverse energy in the event were also considered as inputs to various algorithms. A full list of all variables considered for each approach is given in Table~\ref{tab:allVariables}. All variables were evaluated using the Kolmogorov-Smirnov (KS) test for individual separation power between signal and background. Variables were then sorted in descending order of KS separability. Each algorithm is trained on a subset that minimizes the number of variables while not sacrificing performance.

\begin{table}[ht!]
 \label{tab:allVariables}
\centering
  %\begin{center}
    \begin{tabular}{|c|c|c|c|} % <-- Alignments: 1st column left, 2nd centered, with vertical lines in between
      \hline\hline
      \multirow{2}{*}{Event-Level} & \multicolumn{3}{c|}{Reconstructed}\\
      \cline{2-4}
      & di-Higgs System & Higgs Candidates & Jet Variables \\
      \hline
      $N_{\textrm{jets}}, N_{\textrm{b-tags}}$ & $m_{\textrm{hh}}, p_{\textrm{T}}^{\textrm{hh}}, \eta^{\textrm{hh}}, \phi^{\textrm{hh}},$ & $m_{\textrm{h1}}, p_{\textrm{T}}^{\textrm{h1}}, \eta^{\textrm{h1}}, \phi^{\textrm{h1}},$ & $p_{\textrm{Tj}}^{(i=1-4)}, \eta_{\textrm{Tj}}^{(i=1-4)}, \phi_{\textrm{Tj}}^{(i=1-4)},$ \\
      Missing $E_{\textrm{T}},$ & $\Delta R(\textrm{h1, h2}),$& $m_{\textrm{h2}}, p_{\textrm{T}}^{\textrm{h2}}, \eta^{\textrm{h2}}, \phi^{\textrm{h2}},$ & b-tag$^{(i=1-4)}$\\
      Scalar $H_{\textrm{T}}$ & $\Delta \phi(\textrm{h1, h2})$& $\Delta R(\textrm{h1 jets}), \Delta R(\textrm{h2 jets}),$& \\
      & & $\Delta \phi(\textrm{h1 jets}), \Delta \phi(\textrm{h2 jets})$ & \\
      \hline\hline
    \end{tabular}
\caption{Complete list of all variables considered for analysis in each method. Event-level variables require no di-Higgs reconstruction. Reconstructed variables are defined only after selecting the jet pairs to reconstruct the di-Higgs system in each event. The sub/superscript 'hh' corresponds to the four-body di-Higgs system, the sub/superscript 'h1' corresponds to the leading $p_{\textrm{T}}$ Higgs candidate, and the sub/superscript 'h2' corresponds to the sub-leading $p_{\textrm{T}}$ Higgs candidate. Kinematic for all individual jets chosen by the reconstruction are also available for use in training classification algorithms.}
\end{table}

%% file: supervised_ml.tex
\section{Supervised Learning}
\label{sec:supervised}
Searches for specific signatures or interactions in collider data can be thought of as a classification problem - some known signal process must be identified and separated from some known and well-modeled set of background processes. Any iterative algorithm can then improve its ability to properly identify signal from background by comparing its predicted classifications to the true known classifications and adjusting its internal parameters. This type of approach is known as supervised machine learning, and it is particularly relevant when training models to distinguish between different known processes. We discuss several supervised learning approaches below.

% BDT
\subsection{Boosted Decision Tree}
\input{BDT/bdt.tex}

% BDT
\subsection{Random Forest}
\input{randomForest/rf.tex}

% ff Neural Network
\subsection{Feed Forward Neural Network}
\input{ffNN/nn.tex}

% ff Neural Network
\subsection{Convolutional Neural Network}
\input{CNN/cnn.tex}

%\subsection{Residual Network}

%\subsection{Lorentz Boost Network}
%\input{LBN/lbn.tex}

% EFN network
\subsection{Energy Flow Network}
\input{EFN/efn.tex}

%% file: BDT/bdt.tex
\label{sec:BDT}
Boosted Decision Trees (BDTs) have a long history in high energy physics from enabling the first observation of single top production at the Tevatron~\cite{Abazov:2006gd, Aaltonen:2008sy} to helping in the discovery of the Higgs boson at the LHC~\cite{Aad_2012, Chatrchyan_2012}. A decision tree functions by making a series of sequential cuts (or decisions) that each maximize the separation between signal and background events for a single variable. This separation is maximized by calculating the Gini impurity index

\begin{equation*}
I = 1 - \sum_{i=1}^{J} p_i^2
\end{equation*}
where $i$ runs over the number of classes in a dataset and $p_i$ is the fraction of items labeled as class $i$ in the set after a tree split. By maximizing the purity produced by subsequent cuts, a well-designed decision tree efficiently splits the original dataset into independent sub-populations grouped by class.

Any series of cuts for identifying events will inevitably misclassify some events, and there are many strategies for improving the results. A boosted decision tree attempts to improve the classification by creating a new set of data from the improperly classified events and training a new decision tree on these inputs. Each step of re-training with misclassified events is called a \textit{boost}, and the total prediction for an event is the weighted sum of predictions from the original tree plus the predictions from the boosted trees where each sequential boost receives a smaller weight in the sum.

The BDT trained for di-Higgs detection was built using the xgboost package~\cite{xgboost}. The top seventeen reconstructed and event-level variables ranked by KS separability (see Table~\ref{tab:allVariables} for a summary of variables) were used in training. These included the mass, momentum, and angular separation variables for the di-Higgs system and the Higgs candidate di-jets. Event-level variables (e.g. $N_{\textrm{b-tags}}$, $N_{\textrm{jets}}$) and individual jet momenta were also used in training.

A set of tuneable variables called hyperparameters determine the training behavior and ultimate performance of any machine learning model include BDTs. Of all the hyperparameters that describe the BDT, five were found to have significant impact on the final performance of the model: the multiplicative boost factor described above, the maximum number of decisions allowed per tree (max depth), a minimum separation improvement necessary for a further cut ($\gamma$), a term that reduces large absolute boosted corrections (L1 regularization or $\alpha$), and a term that reduces the impact of large squared corrections in each boost (L2 regularization or $\lambda$). Regularization terms are a common element of many machine learning algorithms and are helpful in avoiding in over-training. The set of hyperparameters that yielded the best significance for the BDT were found to be: boost factor of 0.1, maximum tree depth of 9, $\gamma$ of 1.1, L1 regularization term of 22.1, and an L2 regularization term of 8.3.

\begin{figure}[!h]
\begin{center}
\includegraphics[width=3in]{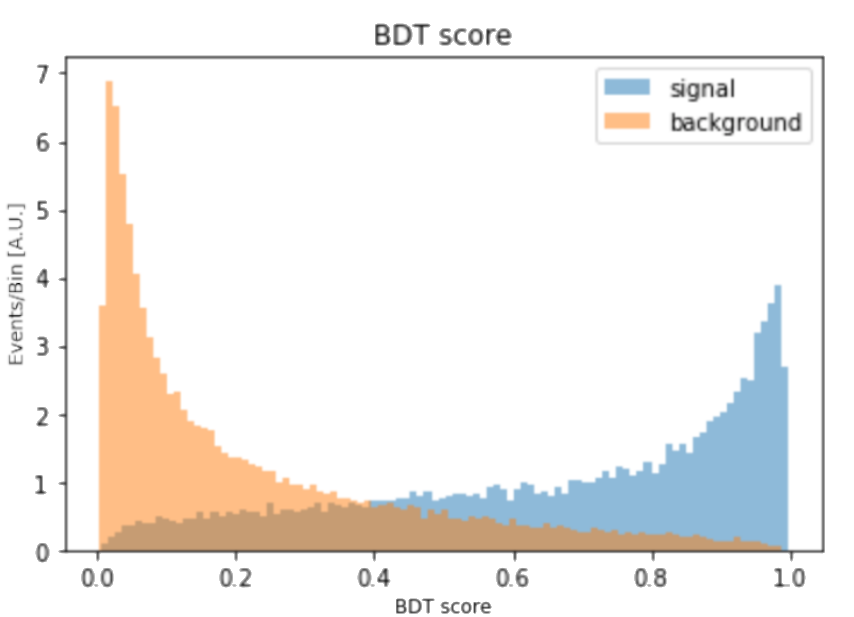}
\caption{Signal predictions of the trained BDT for independent signal and background samples not used for training.}
\label{fig:bdt_pred}
\end{center}
\end{figure}

The predictions from the optimized BDT are shown in Figure~\ref{fig:bdt_pred}. A maximum significance of 1.84$\pm$0.09 was obtained, yielding $986.3 \pm 8.9$ signal events and 2.8$\pm 0.1$ $\cdot$ $10^5$ background events.

%% file: randomForest/rf.tex
\label{sec:RandomForest}
Random Forest algorithms share a similar tree structure with BDTs, but they leverage ensembles of independent decision trees as opposed to iteratively improving the predictions of a single tree using misclassified events. Each tree in a random forest is `grown' using a random sampling of input variables and training events. The randomness of the sampling ensures each tree yields a unique but correlated prediction compared to the other trees in the forest. The class prediction of the forest is the majority vote of the constituent trees. Tuning the hyperparameters of a random forest requires optimizing the number of trees in the forest, the variable sub-sampling used to produce each tree, and the depth of the constituent trees.

The random forest trained for di-Higgs classification uses the reconstruction algorithm that selects di-jet pairs consistent with a Higgs mass hypothesis, and the same reconstructed and event-level variables used for the BDT were used as input to the forest. The random forest was implemented using xgboost~\cite{xgboost}. Five hyperparameters were found to have a meaningful impact on the performance of the model: the number of trees in the forest, the maximum tree depth, a L1 regularization term, the fraction of events used as inputs to each tree (sub-sample fraction), and the fraction of reconstruction variables randomly selected as inputs for each tree (the column sub-sampling rate). Similar to the approach for optimizing the hyperparameters of the BDT, an optimal random forest configuration was obtained by individually varying each significant hyperparameter over a reasonable range and selecting the best performing model.

\begin{figure}[!h] 
\begin{center}
\includegraphics*[width=0.75\textwidth] {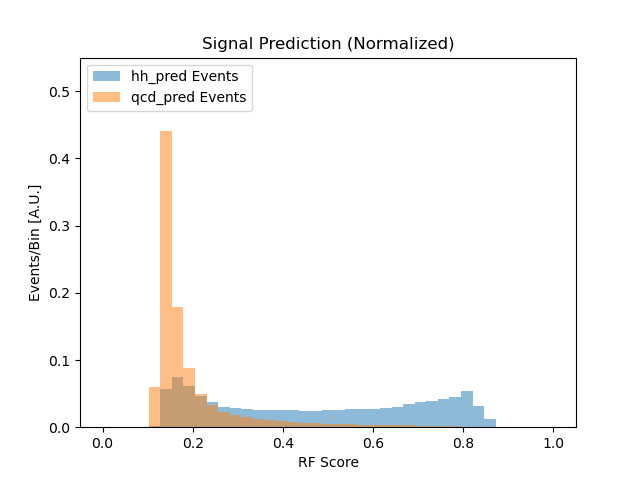}
\caption{Output score on the testing dataset with the fully trained random forest classifier.}
  \label{fig:rf_score}
\end{center}
\end{figure}

The optimal random forest was training with 300 constituent trees, a maximum tree depth of 20, a L1 regularization term of 1.175, a sub-sampling fraction of 0.8, and a column sub-sampling rate of 0.8,. The best significance for the random forest approach was found to be $\sigma$ = 2.44$\pm$0.19 when requiring a prediction score > 0.80. This region yielded an expectation of $544.7 \pm 6.3$ signal events and $5.0 \pm 0.5$ $\cdot$ $10^4$ background events. Distributions of the predictions for signal and background are shown in Figure~\ref{fig:rf_score}.

%% file: ffNN/nn.tex
\label{sec:NN}
Fully connected or feed-forward neural networks (NNs) also have a long history in high energy physics. The fundamental element of any neural network is called a \textit{layer}. Multiple layers are stacked together to produce a final prediction given the input variables from the first layer. This predicted outcome is then evaluated against a known target value. A fully-connected network can have multiple internal (or hidden) layers between the input and output layers, and each hidden layer is composed of a series of activation functions and trainable weights that allow the network to identify and iteratively combine important features of the input space. A function (called the loss function) is chosen to quantify the difference between the model prediction and target values. The loss calculated after a single training iteration is used to adjust the internal network weights in the next training iteration through a process called backpropagation. The model is fully trained once the improvement in the loss between iterations falls beneath some user-defined threshold.

The NN trained for di-Higgs detection was built using the Keras~\cite{chollet2015keras} and Tensorflow~\cite{tensorflow} packages. The top twenty-two most separable reconstructed and event-level variables were used as the input variables for the NN. These included the mass, momentum, and angular separation variables for the di-Higgs system and the Higgs candidate di-jets, momentum and angular information from the individual selected jets, and all event-level variables. Hyperparameter optimization for neural networks involves the additional step of designing the structure of the network: how many hidden layers to use, how many nodes per hidden layer, which activation functions to use in a given layer, using dropout layers which remove random weights in each training iteration to prevent over-fitting, and adding batch normalization layers which stabilize and accelerate network training by re-scaling the neuron weights of the previous layer.

Several models were trained by individually tuning each hyperparameter over a reasonable range in order to produce a final optimized model. The number of hidden layers was similarly optimized by training models with one, two, three, and four hidden layers. No networks with more than four hidden layers were tested in order to keep the number of trainable model parameters below approximately one tenth the number of training events.

\begin{figure}[!h] 
\begin{center}
\includegraphics*[width=0.75\textwidth] {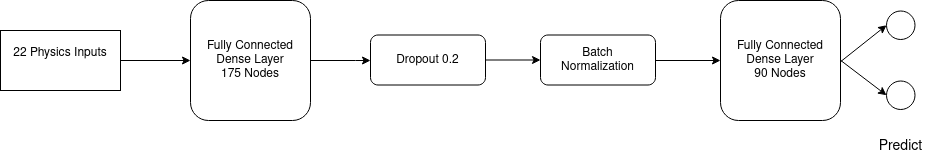}
\caption{Structure of the feed-forward neural network. The input variables are fed through two fully connected dense layers to classify events. One dropout layer and one batch normalization layer help mitigate over-fitting during training.}
  \label{fig:nn}
\end{center}
\end{figure}

The final network structure consisted of the input layer, two hidden layers, and a single-node output layer. The first hidden layer contained 175 nodes with an L2 kernel regularizer ($\lambda$ = $10^{-4}$). The second hidden layer contained 90 nodes with no kernel regularizer. A batch normalization layer and a dropout layer (dropout fraction 0.2) were placed in between the two hidden layers to prevent over-fitting. Both hidden layers used a rectified linear (ReLU) activation function, while the output layer used a sigmoid activation function. The final performance was found to be independent of the choice of activation functions, and they are included here for completeness. The binary cross-entropy loss function

\begin{equation*}
L = -(y\log(p)+(1-y)\log(1-p))
\end{equation*}
was used for the backpropagation. Here $y$ represents the known binary target classifications, $p$ represents the predicted probabilities that a given event is signal or background, and $\log$ is the natural logarithm. A schematic flowchart of the network structure is shown in Figure~\ref{fig:nn}.
%The sequential model was compiled using the ‘adam’ optimizer along with the ‘binary\_crossentropy’ loss method. Finally, the model was fit on the training data along with the validation data for 100 epochs. 

\begin{figure}[!h] 
\begin{center}
   \includegraphics[width = 3in]{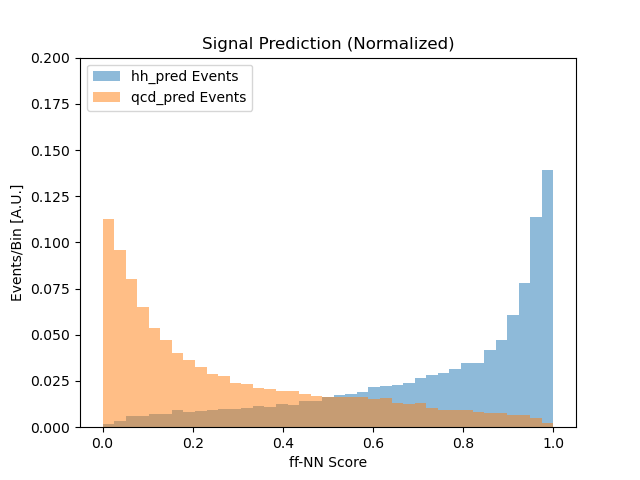}\\
\caption{Final predictions of the feed-forward network for signal and background samples.}
  \label{fig:results_nn}
\end{center}
\end{figure}

The NN was trained for 25 epochs before the minimal loss-improvement threshold was met, and the results are shown in Figure~\ref{fig:results_nn}. The trained model obtained a maximum $\sigma$ = 2.40$\pm$0.08 when considering events with a signal prediction score > 0.94. This phase-space has a signal yield of $1659.9 \pm 12.5$ events and a background yield of $4.8 \pm 0.2$ $\cdot$ $10^5$ events. %477215.3 events.

%% file: CNN/cnn.tex
\label{sec:CNN}
Convolutional Neural Networks (CNNs) are neural networks that predict the content of an input image by using assumptions about the local relationships between neighboring pixels. For this analysis, content prediction is simplified to a general classification of whether the image comes from a di-Higgs or QCD event. The fundamental elements of any convolutional network are convolutional layers and pooling layers. Convolutional layers use filters that perform linear combinations of neighboring pixels within the filter size, and pooling layers aggregate information by grouping neighboring pixels using either their maximum or average values. After some number of these layers, the output is flattened into a one-dimensional vector, and this flattened vector is pushed through a set of feed-forward layers in order to make a final output prediction. 

\begin{figure}[!h] 
\begin{center}
\includegraphics*[width=0.75\textwidth] {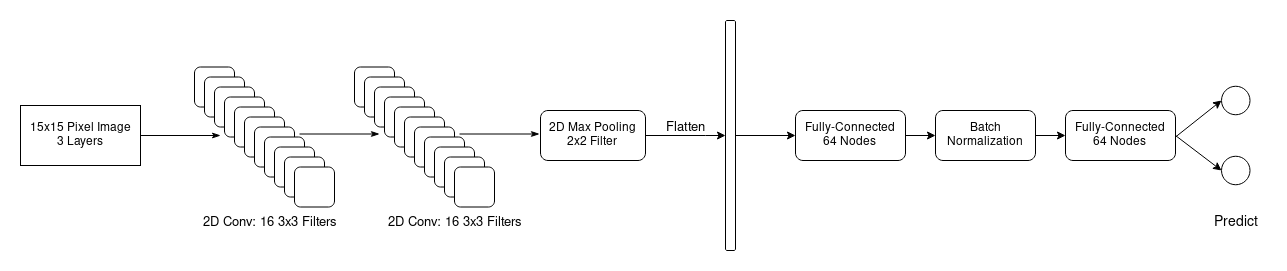}
\caption{Structure of the nominal convolutional neural network. The input images are fed through two convolutional layers and a single max-pooling layer before being flattened into a one-dimensional vector. The flattened vector is then fed through one fully connected layer, a batch normalization layer, and a final fully connected layer before a final prediction is made.}
  \label{fig:cnn_nominal}
\end{center}
\end{figure}

\begin{figure}[!h] 
\begin{center}
\includegraphics*[width=0.75\textwidth] {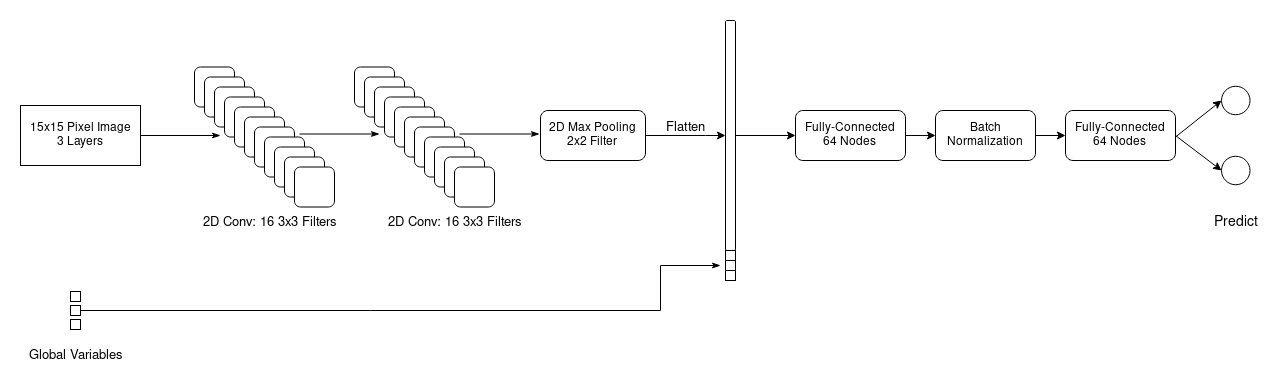}
\caption{Structure of the hybrid convolutional neural network. The input images are fed through two convolutional layers and a single max-pooling layer before being flattened into a one-dimensional vector. Scaled user-specified variables (e.g. $H_{T}$) are then concatenated with the flattened image vector. The concatenated vector is then fed through one fully connected layer, a batch normalization layer, and a final fully connected layer before a final prediction is made.}
  \label{fig:cnn_hybrid}
\end{center}
\end{figure}

Many previous papers have explored the use of convolutional networks trained on low-level quantities (e.g. tracks and calorimeter deposits) for the purposes of object identification~\cite{Alison:2019kud} at colliders. This paper extends the application to event-level identification. Using low-level quantities removes the need to reconstruct higher-level objects like jets or jet pairs; only the detector-level measurements are required for image creation. The performance of four convolutional networks was studied in the context of di-Higgs identification. The first network used a 3-layer image composed of energy/momentum weighted tracks, electromagnetic calorimeter deposits, and hadronic calorimeter deposits. The second network used the same three layers but appended additional global event-level information to the flattened vector after image processing and before the fully connected layers. Figures~\ref{fig:cnn_nominal} and \ref{fig:cnn_hybrid} depict both network structures. The third and fourth networks followed the same pattern as the previous two but with the addition of two image layers corresponding to longitudinal and transverse impact parameter-weighted track information.

\begin{figure}[!ht] 
\begin{center}
  \subfloat[]{\includegraphics[width = 2in]{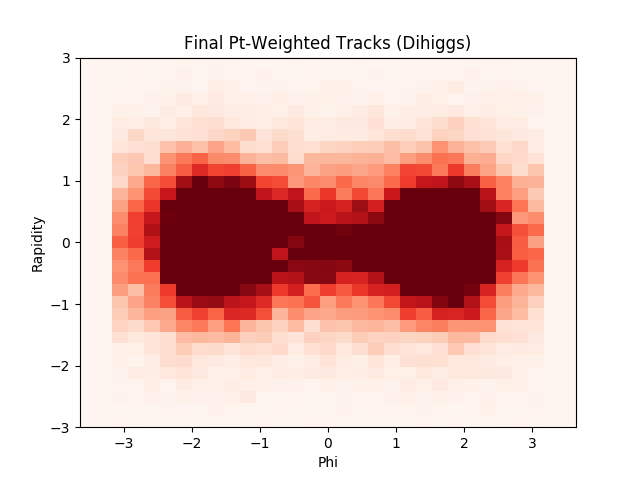}} 
  \subfloat[]{\includegraphics[width = 2in]{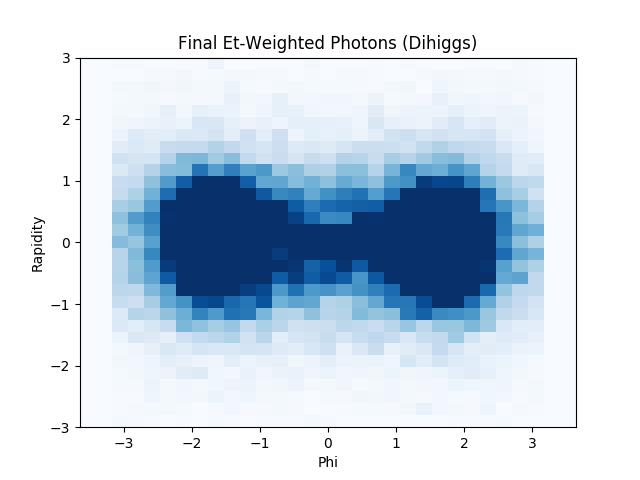}}
  \subfloat[]{\includegraphics[width = 2in]{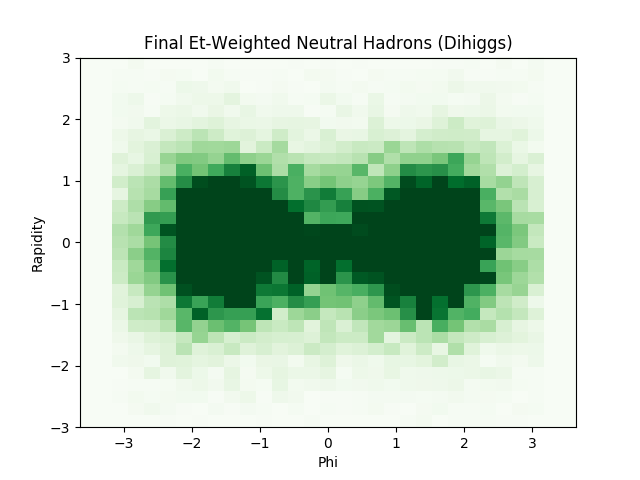}} \\
  \subfloat[]{\includegraphics[width = 2in]{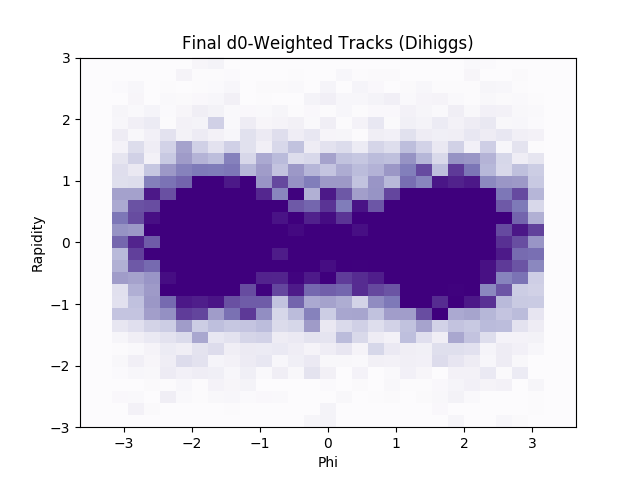}} 
  \subfloat[]{\includegraphics[width = 2in]{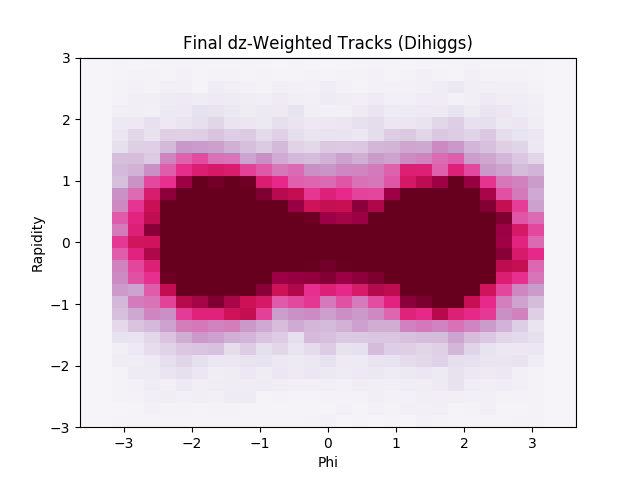}} \\
  \subfloat[]{\includegraphics[width = 2in]{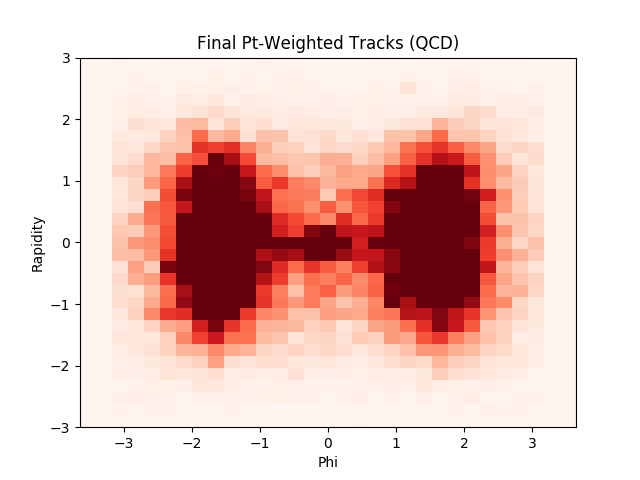}} 
  \subfloat[]{\includegraphics[width = 2in]{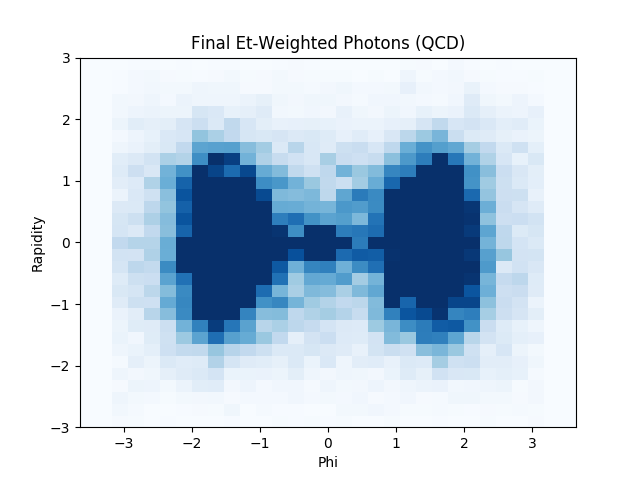}}
  \subfloat[]{\includegraphics[width = 2in]{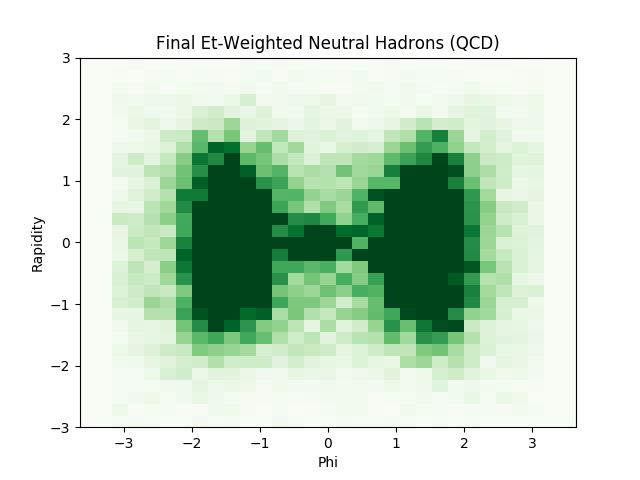}} \\
  \subfloat[]{\includegraphics[width = 2in]{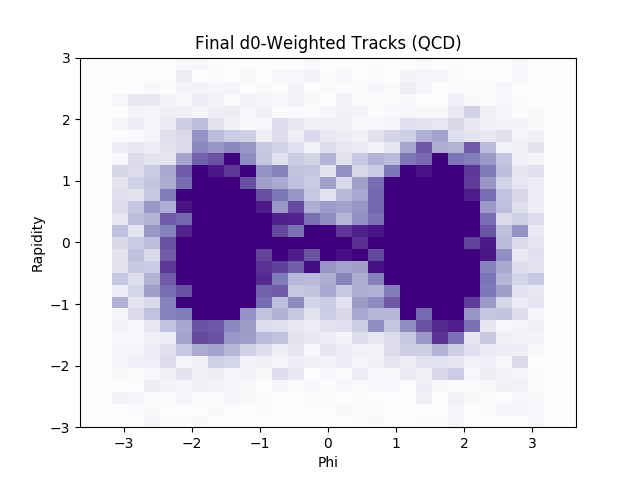}} 
  \subfloat[]{\includegraphics[width = 2in]{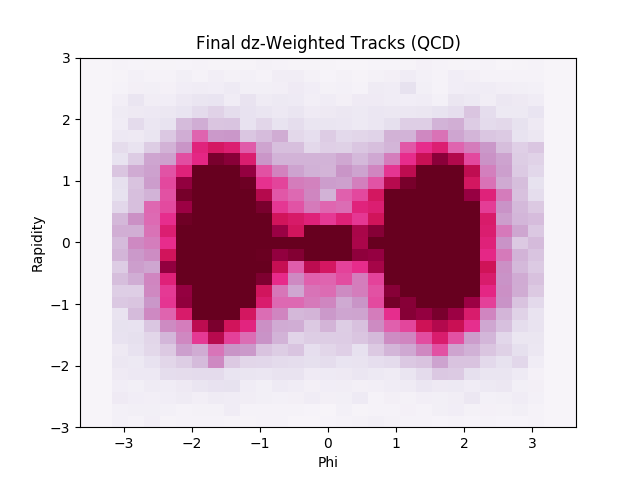}} 
\caption{Average image layers for di-Higgs (a--e) and QCD (f--j) decays. Sub-figures (a) and (f) show $p_{\textrm{T}}$-weighted tracks, (b) and (g) show $E_{\textrm{T}}$-weighted ECAL deposits, (c) and (h) show $E_{\textrm{T}}$-weighted HCAL deposits, (d) and (i) show transverse impact parameter-weighted tracks, (e) and (j) show longitudinal impact parameter-weighted tracks.} %The differences between di-Higgs and QCD layers are apparent.}
\end{center}
\label{fig:cnn_avgImages}
\end{figure}

In order to produce coherent images, the center of mass and the center of momentum for each event are calculated. All constituents are then boosted longitudinally into the center of mass of the event and rotated in phi to the center of momentum. After this pre-processing, each image layer corresponds to a 31x31 pixel grid centered on the total activity in the event. Figure~\ref{fig:cnn_avgImages} shows the average di-Higgs image layers (a--e) and the average QCD image layers (f--j). While the average image layers for each sample closely resemble one another, they do contain different information, and variations are visible.

Importantly, clear differences are observed between the average QCD images and the average di-Higgs images. Each half of the di-Higgs image (split across $\phi$ = 0) is arranged in a roughly circular, isotropic shape due to the spin-0 nature of the Higgs. The QCD images appear balanced because of the effect of the pre-processing, but no similar circular structure is produced. Additionally, the variance of pixel intensities in di-Higgs images is much smaller than the variance in QCD images due to the more balanced kinematics of Higgs pair production compared to QCD processes.

%\begin{figure}[!h] 
%\begin{center}
%  \subfloat[]{\includegraphics[width = 2in]{CNN/figures/images/hh/Tracks_Dihiggs_Final_Pt-Weighted}} 
%  \subfloat[]{\includegraphics[width = 2in]{CNN/figures/images/hh/Photons_Dihiggs_Final_Et-Weighted}}
%  \subfloat[]{\includegraphics[width = 2in]{CNN/figures/images/hh/NeutralHadrons_Dihiggs_Final_Et-Weighted}} \\
%  \subfloat[]{\includegraphics[width = 2in]{CNN/figures/images/hh/Tracks_Dihiggs_Final_d0-Weighted}} 
%  \subfloat[]{\includegraphics[width = 2in]{CNN/figures/images/hh/Tracks_Dihiggs_Final_dz-Weighted}} 
%\caption{Average di-Higgs image showing (a) $p_{\textrm{T}}$-weighted tracks, (b) $E_{\textrm{T}}$-weighted ECAL deposits, (c) $E_{\textrm{T}}$-weighted HCAL deposits, (d) transverse impact parameter-weighted tracks, (e) longitudinal impact parameter-weighted tracks.}
%\end{center}
%\label{fig:cnn_avgDihiggs}
%\end{figure}

\begin{figure}[!h] 
\begin{center}
  \includegraphics[width = 3in]{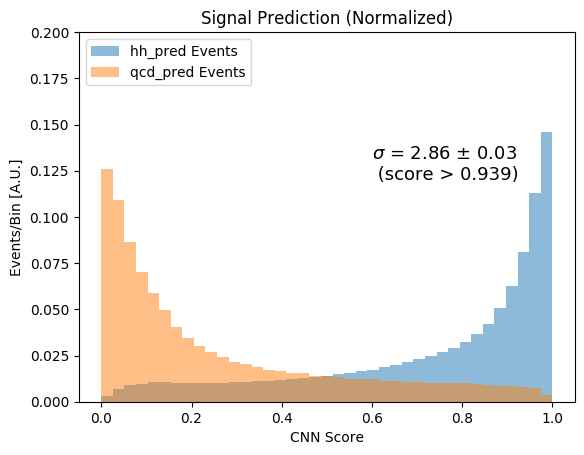}
  %\subfloat[]{\includegraphics[width = 3in]{CNN/figures/5color_0PU_pix31_addHT-nJets-nBTags_2Conv16-16_one2DPool_EqualSamples/compositeImgs_noWeights_80percent_ROC}}\\
\caption{Signal prediction for the 5-color convolutional network with additional high-level inputs. The total area of the signal and background predictions are normalized to unity for easier shape comparison.}
\end{center}
\label{fig:cnn_preds}
\end{figure}

As shown in Figures~\ref{fig:cnn_nominal} and \ref{fig:cnn_hybrid}, the CNN network structure uses two sequential 2D convolutional layers each with 16 3x3 filters, one max-pooling layer with a 2x2 window, a flattening of the outputs, two 64-node fully connected hidden layers, and one output layer for making the final prediction. As previously described, two of the networks append additional high level variables (scalar sum of transverse hadronic energy, number of jets, and number of $b$-tags) after the flattening and before the image information is fed through the fully connected layers. The optimal significance for each network is shown in Table~\ref{tab:cnnResults}. The best results were obtained using the 5-color network with additional high-level inputs, and the final predictions for this configuration are shown in Figure~\ref{fig:cnn_preds}. A best significance of 2.86$\pm$0.03 was found for a prediction cut $>$ 0.94 with a signal yield of $1.00\pm 0.05$ $\cdot$ $10^4$ events and a background yield of $1.32\pm 0.01$ $\cdot$ $10^7$ events.

\begin{table}[h!]
\label{tab:cnnResults}
  \begin{center}
    \begin{tabular}{|l|c|c|} % <-- Alignments: 1st column left, 2nd middle and 3rd right, with vertical lines in between
      \hline\hline
      \textbf{Method} & Best $\sigma$ & AUC \\
      \hline
      Tracks+HCAL+ECAL & 1.77 $\pm$ 0.01 & 0.818 \\
      Tracks+HCAL+ECAL + high-level & 2.12 $\pm$ 0.01 & 0.846 \\
      Tracks+HCAL+ECAL+D0+DZ & 2.45 $\pm$ 0.02 & 0.863 \\
      Tracks+HCAL+ECAL+D0+DZ + high-level & 2.86 $\pm$ 0.03 & 0.882 \\

      \hline\hline
    \end{tabular}
    \caption{Normalized to full HL-LHC dataset of 3000 fb$^{-1}$}
  \end{center}
\end{table}

%% file: EFN/efn.tex
\begin{figure}[ht!]
\centering
\includegraphics[scale=0.5]{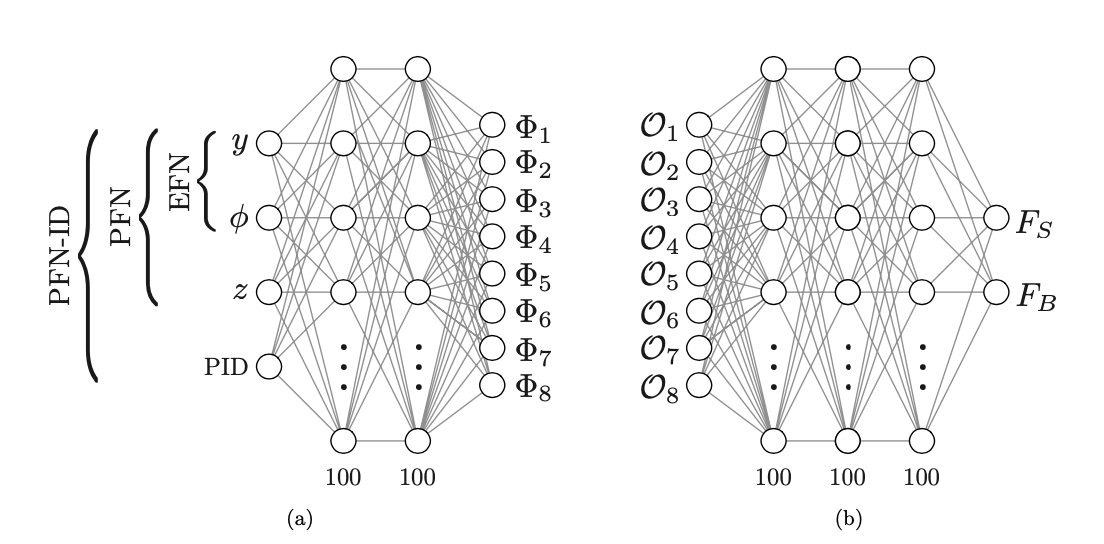}
\caption{Example of a complete EFN/PFN network. Network (a) uses jet constituent information as input and outputs a set of latent variables, $\Phi$. Network (b) produces linear combinations, $\mathcal{O}$, of the latent variables which are then fed through several fully connected layers to make a final prediction.}
\label{fig:EFNArch}
\end{figure}

\label{sec:EFN}
Energy Flow Networks (EFN) and Particle Flow Networks (PFN) are neural networks that also operate with basic jet constituent information as input rather than reconstructed jets and multi-jet composites~\cite{Komiske:2018cqr}. The EFN structure takes only the rapidity, ${y}$, and azimuthal angle, ${\phi}$, of jet constituents as input, while the PFN takes the rapidity, azimuthal angle, and transverse momentum, $p_{T}$, of jet constituents as input. Both the EFN and PFN are two-component networks, and their internal structures are shown in Figure \ref{fig:EFNArch}.

Network (a) takes jet constituent information as input and uses a set of fully connected neural network layers to produce a set of latent features, \textbf{$\Phi$}. Linear combinations of the latent variables, \textbf{$\mathcal{O}$} are then used as the input for network (b) which uses another set of fully connected layers to produce final predictions. The EFN and PFN used for di-Higgs classification use 200 nodes for each hidden layer in network (a), 256 nodes for the latent space dimension, and 300 nodes for each hidden layer in network (b). 

The EFN/PFN networks were trained using four separate categories split by number of jets and number of $b$-tags to test the network's dependence on higher-level jet information. Independent networks were trained using: all events, only events with $\geq$4 jets, only events with $\geq$4 jets and =2 $b$-tags, and only events with $\geq$4 jets and $\geq$4 $b$-tags. In each configuration, the number of signal and background events were adjusted to maintain an equal proportion of each population in the training sample. L2 regularization and dropout layers were added to minimize over-fitting. The results obtained from each EFN configuration are shown in Table~\ref{EFNtab}. The results of each PFN configuration are shown in Table~\ref{PFNtab}.

\begin{table}[ht!]
\centering
  %\begin{center}
    \begin{tabular}{|l|c|c|c|} % <-- Alignments: 1st column left, 2nd middle and 3rd right, with vertical lines in between
      \hline\hline
      \multirow{2}{*}{\textbf{Category}} & \multicolumn{3}{c|}{0PU}\\
      \cline{2-4}
      & Best $\sigma$ & \textbf{N$_{\mathrm{Signal}}$} & \textbf{N$_{\mathrm{Background}}$} \\
      \hline
      All Events & $1.407 \pm 0.006$ & $1.89\pm 0.01 \cdot 10^4$ & $1.80\pm 0.01 \cdot 10^8$ \\
      4Jets & $1.363 \pm 0.006$ & $1.63\pm 0.01 \cdot 10^4$ & $1.43\pm 0.01 \cdot 10^8$ \\
      4Jets 2BTags & $1.343 \pm 0.006$ & $1.33\pm 0.01 \cdot 10^4$ & $9.95\pm 0.06 \cdot 10^7$ \\
      4Jets 4BTags & $0.867 \pm 0.008$ & $3468.7\pm 21.1 $ & $1.60\pm 0.01 \cdot 10^7$ \\
      \hline\hline
    \end{tabular}
    \caption{Results of the optimal EFN performance across categories. Event yields normalized to the full HL-LHC dataset of 3000 fb$^{-1}$. The provided errors take into account the Monte Carlo statistical uncertainty and the uncertainty on the Madgraph generated cross-section.}
  %\end{center}
\label{EFNtab}
\end{table}

\begin{table}[ht!]
\centering
  %\begin{center}
    \begin{tabular}{|l|c|c|c|} % <-- Alignments: 1st column left, 2nd middle and 3rd right, with vertical lines in between
      \hline\hline
      \multirow{2}{*}{\textbf{Category}} & \multicolumn{3}{c|}{0PU}\\
      \cline{2-4}
      & Best $\sigma$ & \textbf{N$_{\mathrm{Signal}}$} & \textbf{N$_{\mathrm{Background}}$} \\
      \hline
      All Events & $1.618 \pm 0.008$ & $1.79\pm 0.01 \cdot 10^4$ & $1.21\pm 0.01 \cdot 10^8$ \\
      4Jets & $1.580 \pm 0.008$ & $1.32\pm 0.01 \cdot 10^4$ & $7.00\pm 0.04 \cdot 10^7$ \\
      4Jets 2BTags & $1.574 \pm 0.009$ & $1.32 \pm 0.01$ $\cdot$ $10^4$ & $4.85\pm 0.03 \cdot 10^7$ \\
      4Jets 4BTags & $0.903 \pm 0.009$ & $3297.3 \pm 20.4$ & $1.33\pm 0.01 \cdot 10^7$ \\
      \hline\hline
    \end{tabular}
    \caption{Results of the optimal PFN performance across categories. Event yields normalized to the full HL-LHC dataset of 3000 fb$^{-1}$. The provided errors take into account the Monte Carlo statistical uncertainty and the uncertainty on the Madgraph generated cross-section.}
  %\end{center}
\label{PFNtab}
\end{table}

Both networks performed best when trained over all events without any cuts on the number of jets or $b$-tags. The EFN obtained a highest significance of 1.41$\pm$0.01, and the PFN obtained a highest significance of 1.62$\pm$0.01.

%% file: semi-supervised_ml.tex
\section{Semi-Supervised Learning}
\label{sec:semisupervised}
While it makes sense to treat searches for new physics or rare signatures as a supervised classification problem, an alternative approach is to let an algorithm learn intrinsic features from an unlabeled dataset and then evaluate whether this self-learned information can be used to identify sub-samples within a complete data set. The process of learning features from unlabeled datasets is called unsupervised learning. Unsupervised machine learning can be transformed into semi-supervised learning when the results of unsupervised learning are evaluated using known classification information.

\subsection{$k$-Means Clustering}
\input{BDT/kmeans.tex}

\subsection{Autoencoder}
\input{AE/ae.tex}

%% file: BDT/kmeans.tex
\label{sec:kmeans}
K-means clustering is an unsupervised learning algorithm that finds natural unlabeled groupings in a scaled phase-space describing the input dataset. Scaled data is created by compressing or stretching each variable of the dataset to the same maximum and minimum values. This ensures each variable is treated equally when calculated the distances between neighboring points. Minimum values of 0 and maximum values of 1 were chosen for all variables in this analysis.

The k-means approach creates clusters by randomly seeding a user-specified number of cluster centroids in the scaled phase-space. Each point is associated to the closest nearby centroid, and the group of points belonging to a single centroid is defined as a cluster. The centroid positions are updated over several iterations by minimizing the ensemble sum of the squared distance between a centroid and all points associated to that cluster. As the centroids move, points can be associated to different clusters eventually obtaining a set of clusters defined by locally dense groupings of similar points. Combining unsupervised clustering with the supervised structure of a BDT converts the unsupervised approach into a semi-supervised algorithm whose performance can be compared to other supervised methods.

The number of clusters, $k$, to fit is a user-defined hyperparameter, and the number of clusters used to describe a dataset must be carefully chosen. Using too few clusters risks not being able to identify meaningfully different populations in the input dataset. Using too many clusters risks losing predictive power by splitting coherent sets of points into arbitrary groupings. An optimal $k$ is found by scanning several values, calculating the total distance between all points and their associated centroid, and selecting a $k$ that minimizes this distance without asymptotically approaching zero.

The optimal $k$ for the di-Higgs data was found lie around 20, and three different clusterings ($k$= 15, 20, 40) were tested for completeness. The scenarios with 15 and 40 clusters test the effects of under-clustering and over-clustering respectively. Two sets of clustered data were used as input to the nominal BDT to test the performance of this semi-supervised approach. The first set used all reconstructed kinematic variables as the input to the $k$-means clustering stage before training the BDT. The second set was produced by performing a principal component analysis (PCA) decomposition on the nominal kinematic inputs before passing through the clustering step and the BDT. PCA is a technique for finding an orthogonal basis of the input data that minimizes the variance along each new axis. No transformation was found to improve the performance of the nominal configuration, and the results are shown in Table~\ref{tab:bdtPCACluster}.

\begin{table}[h!]
\label{tab:bdtPCACluster}
\begin{center}
  %\hskip-4.0cm
    \begin{tabular}{|l|c|c|c|} % <-- Alignments: 1st column left, 2nd middle and 3rd right, with vertical lines in between
      \hline\hline
      \textbf{Method} & $\sigma$ & N$_{\textrm{sig}}$ & N$_{\textrm{bkg}}$ \\
      \hline
      Nominal BDT & 1.84 $\pm$ 0.09       & $986.3\pm 8.9$   & $2.9\pm 0.1 \cdot 10^5$ \\
      15 Clusters & 1.29 $\pm$ 0.02       & $2100.2\pm 14.6$ & $2.7\pm 0.1 \cdot 10^6$ \\
      15 Clusters + PCA & 1.25 $\pm$ 0.02 & $2189.5\pm 15.1$ & $3.1\pm 0.1 \cdot 10^6$ \\         
      20 Clusters & 1.30 $\pm$ 0.02       & $2260.6\pm 15.4$ & $3.0\pm 0.1 \cdot 10^6$ \\
      20 Clusters + PCA & 1.27 $\pm$ 0.03 & $1756.4\pm 12.9$ & $1.9\pm 0.1 \cdot 10^6$ \\         
      40 Clusters & 1.44 $\pm$ 0.03       & $1704.6\pm 12.7$ & $1.4\pm 0.1 \cdot 10^6$ \\
      40 Clusters + PCA & 1.34 $\pm$ 0.02 & $2144.5\pm 14.8$ & $2.0\pm 0.1 \cdot 10^6$ \\         
      \hline\hline
    \end{tabular}
    \caption{Significance and yields showing BDT performance when using the nominal kinematic inputs, clustered kinematic inputs, and clustered inputs from a PCA decomposition. All yields are normalized to full HL-LHC dataset of 3000 fb$^{-1}$.}
    \end{center}
\end{table}

%% file: AE/ae.tex
\label{sec:AE}

An autoencoder (AE) is an unsupervised machine learning architecture used for detecting anomalies that differ significantly from the data used to train the network. The structure of the AE compresses the input information into a lower-dimensional representation called the latent space. This compression `encodes' the most important features of the training data into the latent space, while the second half of the network `decodes' the latent space back into a representation approaching the original inputs.

This construction fundamentally changes the meaning of the loss calculation-- rather than computing the loss between a prediction and a target, the AE loss is a measure of how well the network reproduces the original inputs after encoding and decoding. Inputs that differ significantly from the data used to train the AE will not be properly reconstructed, and anomalies can be identified by selecting events with large losses. Training with Monte Carlo simulations allows for a semi-supervised cross-check on AE performance since the classes of training and testing samples are known in advance.

Because AE anomaly detection relies on a well-modeled understanding of background processes, the network was trained using only QCD events. Additional models were trained by substituting the pure-QCD training sample with training samples consisting of mixtures of QCD and di-Higgs events to test the stability of the method against signal contamination. No significant deterioration was observed for reasonable levels of contamination. The AE used for di-Higgs detection was built using the Keras package~\cite{chollet2015keras} and consists of an input layer, a single hidden layer, and an output layer. Eleven reconstructed variables (di-Higgs and di-jet Higgs candidate masses, momenta, and angular variables) were selected for use in the AE. 

\begin{figure}[!h] 
\begin{center}
\includegraphics*[width=3.5in] {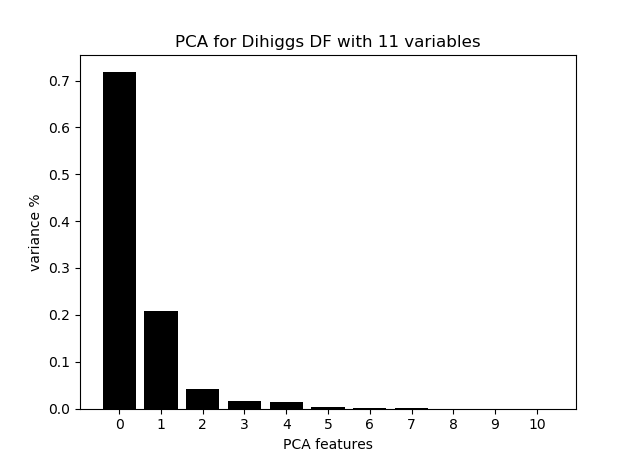}
\caption{PCA performed on the selected eleven kinematic inputs. The x-axis indicates the number of PCA features, and the y-axis indicates the variance. Choosing the optimal size for the latent space requires identifying the point of diminishing variance return.}
  \label{fig:ae_pca}
\end{center}
\end{figure}

The output layer is a mirror of the input layer and therefore has 11 nodes. Similar to the $k$-means clustering, the size of the latent space should be large enough to model the intrinsic features of the input data without being so large that the latent representation is able to learn features drawn arbitrarily from the input space. A PCA analysis (shown in Figure~\ref{fig:ae_pca}) was used to determine that a latent space of three nodes was able to capture over 95\% of the input variance. The hidden layer and output layer use ReLU and sigmoid activation functions respectively. An L2 regularization term was added to the hidden layer to avoid over-fitting. %Afterwards the model was compiled using the ‘adam’ optimizer with the mean-squared error loss. Finally, the model was fit on the training data along with the validation data for 10000 epochs. The ‘EarlyStopping’ function was used to shorten training time by stopping training after the validation accuracy didn’t increase by 0.001 for 50 epochs. The batch size during training was set to 1024 in order to both decrease training time and prevent overfitting.
Because training an autoencoder is an unsupervised and unlabeled process, there is no prediction of whether a given event is signal or background. Since di-Higgs events should be relatively anomalous compared to the QCD training set, signal events should have a relatively larger average loss value. Cutting on the loss function allows for a significance to be calculated for comparison with other methods.

\begin{figure}[!h] 
  \begin{center}
    \subfloat[]{\includegraphics[width = 3in]{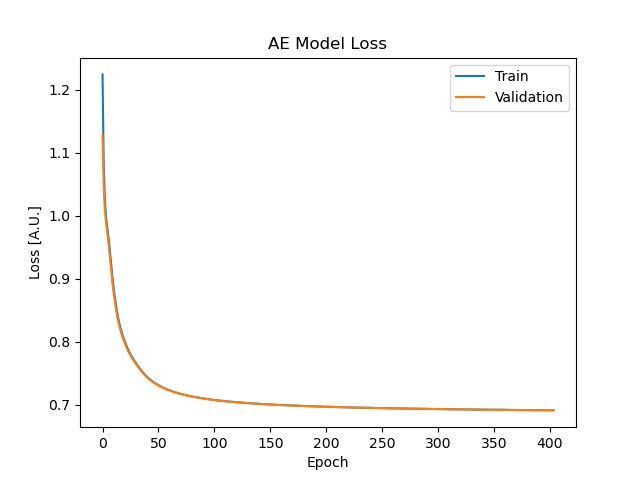}} 
    \subfloat[]{\includegraphics[width = 3in]{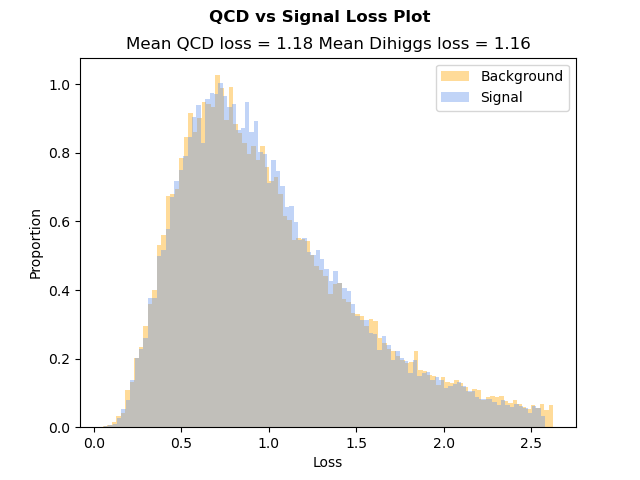}}\\
    \caption{(Left) The loss of the AE during QCD training/reconstruction converged after 700 epochs. (Right) The loss distribution generated by the AE when being tested on QCD and di-Higgs event data separately.}    
  \label{fig:ae_trainPredLoss}
\end{center}
\end{figure}

Training for several hundred epochs leads the model to converge, and it reaches an asymptotic ensemble loss value near 0.7 (see Figure~\ref{fig:ae_trainPredLoss}). A best significance of $\sigma$ = $0.81\pm 0.01$ was obtained for events with individual reconstructed loss scores greater than 0.05. The significance obtained for lower thresholds allowed for slightly more background while the significance obtained for higher loss thresholds rejected too many signal events. Still, this optimal point is somewhat misleading since the signal and background loss distributions have little separation. The highest significance result effectively is a cut that keeps nearly all events. This suggests that the kinematic inputs used in training do not significantly differ between signal and background processes after the latent space compression.

%\begin{figure}[!h] 
%\begin{center}
%\includegraphics*[width=0.75\textwidth] {AE/figures/ae_finalSignalPredictions_qcdTrain}
%\caption{Signal predictions made by the AE based on the loss distributions from Fig. 3. The $S/\sqrt{B}$ best cut was placed near 0.1, indicating that the AE was not able to sufficiently distinguish di-Higgs da%ta from QCD data.}
%  \label{fig:ae_signalPred}
%\end{center}
%\end{figure}

%% file: results.tex
\section{Results}
\label{sec:results}

The methods covered in this paper are by no means an exhaustive review of the ML landscape available to high energy physics. Still, a wide range of techniques and philosophies are covered. Table~\ref{tab:summary} provides a summary of the methods described in the previous sections. The raw yields after selection and reconstruction cuts (depending on the method) have been weighted by the effective cross-sections listed in Table~\ref{tab:samples} and scaled to 3000 fb$^{-1}$. The result from a traditional 1-D sequential cut technique is shown for comparison though the details were not discussed in the previous sections. Clear gains in sensitivity compared to this baseline are apparent for many of the ML models tested in this review.

\begin{table}[h!]
\label{tab:summary}
  \begin{center}
  \begin{tabular}{|l|c|c|c|} % <-- Alignments: 1st column left, 2nd middle and 3rd right, with vertical lines in between
      \hline\hline
      \multirow{2}{*}{\textbf{Method}} & \multicolumn{3}{c|}{0PU} \\
      \cline{2-4}
      & Best $\sigma$ & \textbf{N$_{\mathrm{Signal}}$} & \textbf{N$_{\mathrm{Background}}$} \\
      \hline
      Autoencoder           & $0.81 \pm 0.01$ & $5840.8 \pm 31.8$ & $5.2\pm 0.3$ $\cdot$ $10^7$ \\
      1D-Rectangular Cuts   & $0.82 \pm 0.02$ & $3621.0 \pm 21.8$  & $2.0 \pm 0.2$ $\cdot$ $10^7$ \\
      k-Means Clustering    & $1.44 \pm 0.02$ & $1703.6 \pm 12.7$ & $1.39\pm 0.03$  $\cdot$ $10^6$ \\
      Particle Flow Network & $1.62 \pm 0.01$ & $1.78 \pm 0.08$ $\cdot$ $10^4$ & $1.21\pm 0.01$ $\cdot$ $10^8$ \\
      Boosted Decision Tree & $1.84 \pm 0.09$ & $986.3 \pm 8.9$  & $2.8 \pm 0.1$ $\cdot$ $10^5$ \\
      %Lorentz Boost Network & 1.87 $\pm$ 0.08 & 1123.3 & 3.6 $\cdot$ $10^5$ \\
      Feed-Forward NN       & $2.40 \pm 0.08$ & $1659.9 \pm 12.9$  & $4.8 \pm 0.2$ $\cdot$ $10^5$ \\
      Random Forest         & $2.44 \pm 0.19$ & $544.7 \pm 6.3$ & $5.0 \pm 0.5$ $\cdot$ $10^4$ \\
      Convolutional NN      & $2.85 \pm 0.02$ & $1.00\pm 0.05$ $\cdot$ $10^4$ & $1.32\pm 0.01$ $\cdot$ $10^7$ \\
      \hline\hline
    \end{tabular}
    \caption{Comparison of method significance and signal/background yields normalized to full HL-LHC dataset of 3000 fb$^{-1}$. The provided errors take into account the Monte Carlo statistical uncertainty and the uncertainty on the Madgraph generated cross-section.}
  \end{center}
\end{table}

An important caveat to keep in mind is that all results discussed here were determined in conditions with zero pileup. In higher pileup environments like those expected at the HL-LHC, reconstruction algorithms see serious reductions in correct combinatoric matching. This effect will certainly degrade the expected performance of techniques that rely on explicit event reconstruction. Methods that do not rely on event reconstruction (CNN, PFN) might be more robust to these effects, and this should be studied in further work. The unsupervised AE technique performed the worst among all the methods tested, but this is likely a reflection of the fact that model-specific methods often outperform model-unspecific methods when evaluated on the model used in training.

%% file: conclusions.tex
\section{Conclusions}
\label{sec:conclusions}
Measuring the rate of di-Higgs production will be a problem facing the high energy physics community through the end of the HL-LHC era. The techniques explored in this paper show the power of machine learning techniques in identifying di-Higgs signals amid overwhelming QCD backgrounds. The significance obtained for many of these methods is impressive given the simplicity of the evaluation metric. This bodes well for both current and future measurements of Higgs pair production at the LHC.

%% file: dihiggs_draft_jhep.bbl
\providecommand{\href}[2]{#2}\begingroup\raggedright\begin{thebibliography}{10}

\bibitem{Abazov:2006gd}
{\scshape D0} collaboration, \emph{{Evidence for Production of Single Top
  Quarks and First Direct Measurement of |Vtb|}},
  \href{https://doi.org/10.1103/PhysRevLett.98.181802}{\emph{Phys. Rev. Lett.}
  {\bfseries 98} (2007) 181802}
  [\href{https://arxiv.org/abs/hep-ex/0612052}{{\ttfamily hep-ex/0612052}}].

\bibitem{Aaltonen:2008sy}
{\scshape CDF} collaboration, \emph{{{Measurement of the Single Top Quark
  Production Cross Section at CDF}}},
  \href{https://doi.org/10.1103/PhysRevLett.101.252001}{\emph{Phys. Rev. Lett.}
  {\bfseries 101} (2008) 252001}
  [\href{https://arxiv.org/abs/0809.2581}{{\ttfamily 0809.2581}}].

\bibitem{albertsson2018machine}
K.A.~et~al., \emph{{Machine Learning in High Energy Physics Community White
  Paper}},  2018.

\bibitem{deFlorian:2016spz}
{\scshape LHC Higgs Cross Section Working Group} collaboration, \emph{{Handbook
  of {LHC Higgs Cross Sections}: 4. {Deciphering} the {Nature} of the {Higgs}
  {Sector}}},  \href{https://arxiv.org/abs/1610.07922}{{\ttfamily 1610.07922}}.

\bibitem{Brun:1997pa}
R.~Brun and F.~Rademakers, \emph{{ROOT: An Object Oriented Data Analysis
  Framework}}, \href{https://doi.org/10.1016/S0168-9002(97)00048-X}{\emph{Nucl.
  Instrum. Meth. A} {\bfseries 389} (1997) 81}.

\bibitem{Alwall:2014hca}
J.~Alwall, R.~Frederix, S.~Frixione, V.~Hirschi, F.~Maltoni, O.~Mattelaer
  et~al., \emph{{The Automated Computation of Tree-Level and Next-to-Leading
  Order Differential Cross Sections, and their Matching to Parton Shower
  Simulations}}, \href{https://doi.org/10.1007/JHEP07(2014)079}{\emph{JHEP}
  {\bfseries 07} (2014) 079} [\href{https://arxiv.org/abs/1405.0301}{{\ttfamily
  1405.0301}}].

\bibitem{Sj_strand_2015}
T.~Sjöstrand, S.~Ask, J.R.~Christiansen, R.~Corke, N.~Desai, P.~Ilten et~al.,
  \emph{{An Introduction to PYTHIA 8.2}},
  \href{https://doi.org/10.1016/j.cpc.2015.01.024}{\emph{Computer Physics
  Communications} {\bfseries 191} (2015) 159–177}.

\bibitem{de_Favereau_2014}
J.~de~Favereau, C.~Delaere, P.~Demin, A.~Giammanco, V.~Lemaître, A.~Mertens
  et~al., \emph{{DELPHES 3: A Modular Framework for Fast Simulation of a
  Generic Collider Experiment}},
  \href{https://doi.org/10.1007/jhep02(2014)057}{\emph{Journal of High Energy
  Physics} {\bfseries 2014} (2014) }.

\bibitem{github}
B.~Tannenwald, A.~Li, A.~Cuddeback, R.~Parvatam and C.~Thompson,
  ``{dihiggsMLProject}.''
  \url{https://github.com/neu-physics/dihiggsMLProject}, 2020.

\bibitem{Aad_2012}
G.~Aad, T.~Abajyan, B.~Abbott, J.~Abdallah, S.~Abdel~Khalek, A.~Abdelalim
  et~al., \emph{{Observation of a New Particle in the Search for the Standard
  Model Higgs boson with the ATLAS detector at the LHC}},
  \href{https://doi.org/10.1016/j.physletb.2012.08.020}{\emph{Physics Letters
  B} {\bfseries 716} (2012) 1–29}.

\bibitem{Chatrchyan_2012}
S.~Chatrchyan, V.~Khachatryan, A.~Sirunyan, A.~Tumasyan, W.~Adam, E.~Aguilo
  et~al., \emph{{Observation of a New Boson at a Mass of 125 GeV with the CMS
  experiment at the LHC}},
  \href{https://doi.org/10.1016/j.physletb.2012.08.021}{\emph{Physics Letters
  B} {\bfseries 716} (2012) 30–61}.

\bibitem{xgboost}
T.~Chen and C.~Guestrin, \emph{{XGBoost: {A} Scalable Tree Boosting System}},
  {\emph{CoRR} {\bfseries abs/1603.02754} (2016) }
  [\href{https://arxiv.org/abs/1603.02754}{{\ttfamily 1603.02754}}].

\bibitem{chollet2015keras}
F.~Chollet, ``Keras.'' \url{https://github.com/fchollet/keras}, 2015.

\bibitem{tensorflow}
M.A.~et~al., \emph{{TensorFlow: Large-Scale Machine Learning on Heterogeneous
  Distributed Systems}}, {\emph{CoRR} {\bfseries abs/1603.04467} (2016) }
  [\href{https://arxiv.org/abs/1603.04467}{{\ttfamily 1603.04467}}].

\bibitem{Alison:2019kud}
J.~Alison, S.~An, P.~Bryant, B.~Burkle, S.~Gleyzer, M.~Narain et~al.,
  \emph{{End-to-end Particle and Event Identification at the Large Hadron
  Collider with CMS Open Data}},  in \emph{{Meeting of the Division of
  Particles and Fields of the American Physical Society}}, 10, 2019
  [\href{https://arxiv.org/abs/1910.07029}{{\ttfamily 1910.07029}}].

\bibitem{Komiske:2018cqr}
P.T.~Komiske, E.M.~Metodiev and J.~Thaler, \emph{{Energy Flow Networks: Deep
  Sets for Particle Jets}},
  \href{https://doi.org/10.1007/JHEP01(2019)121}{\emph{JHEP} {\bfseries 01}
  (2019) 121} [\href{https://arxiv.org/abs/1810.05165}{{\ttfamily
  1810.05165}}].

\end{thebibliography}\endgroup
